# EU law and emotion data

Andreas Häuselmann
*eLaw Center for Law and Digital Technologies*
*Leiden University*
Leiden, The Netherlands
a.n.hauselmann@law.leidenuniv.nl
ORCID 0009-0005-3905-1651

Alan M. Sears
*eLaw Center for Law and Digital Technologies*
*Leiden University*
Leiden, The Netherlands
a.m.sears@law.leidenuniv.nl
ORCID 0000-0001-7610-9000

Eduard Fosch-Villaronga
*eLaw Center for Law and Digital Technologies*
*Leiden University*
Leiden, The Netherlands
e.fosch.villaronga@law.leidenuniv.nl
ORCID 0000-0002-8325-5871

Lex Zard
*eLaw Center for Law and Digital Technologies*
*Leiden University*
Leiden, The Netherlands
a.zardiashvili@law.leidenuniv.nl
ORCID 0000-0002-7190-9292

*Abstract—* This article sheds light on the intricate legal implications and challenges surrounding emotion data processing within the EU data protection framework. Despite the sensitive nature of emotion data, the GDPR does not explicitly categorize it as special data, resulting in a lack of comprehensive protection. This legal ambiguity poses significant obstacles for affective computing system developers as they struggle to comply with the GDPR's requirements and ensure ethical practices. The article also discusses the nuances of different approaches in affective computing and their relevance to the GDPR and the introduction of biometric-based data in the AI Act proposal. Moreover, it highlights potential conflicts with GDPR principles, such as fairness and accuracy, and the limitations of the AI Act in addressing specific harmful uses of emotion data. Additionally, the article outlines the new obligations and transparency requirements introduced by the DSA for online platforms utilizing emotion data, making it crucial for the affective computing community to be well-informed in order to adhere to the regulations, maintain ethical and legal standards, and protect users' fundamental rights while developing emotion-sensing technologies for the EU market.

## I. Introduction

Emotions are inherent to humanity and play a significant role in human behavior, communication, and interaction [1]. Until recently, emotions could not be captured and processed automatedly; they were reserved for each individual and the people with whom they wanted to share them. With advancements in computing that relate to, arise from, or influence emotions [2] ('affective computing'), something once unimaginable has become a reality. Affective computing (AC) systems can now capture emotional data.

Emotion data is information about a person's inner emotional state, such as their subjective response to a thing, person, or situation. It can encompass quantitative and qualitative data, such as physiological measurements, facial expressions, speech, and self-reports of feelings captured through technical means [3-4]. Thanks to AC, emotions became machine-readable. Emotion data detected by AC systems increasingly supports and informs ulterior decision-making processes in several fields, including marketing, healthcare, border control, and education, among others; it is often claimed that such use can provide valuable insight into how people feel and respond to different situations [5].

Emotions are complex—a person may not be able to express in words how they feel, and others may or may not understand their meaning. As one can imagine, capturing inner, subjective states through 'objective' technical means may be a difficult task that can lead to errors, as in the case of lie detectors [6] or gender classifier systems [7]. If the contested relationship between several physiological states and several emotions is assumed to connect with potential user behaviors directly, such inferences could lead to disastrous consequences depending on the application context, such as border control [8].

Centuries ago, studies began to explore the biological and evolutionary underpinnings of emotions, including their influence on decision-making, memory, and learning [9-11]. Despite this, there is little understanding of how the law regulates emotional data and how this impacts the field of *affective computing*. With this article, we want to shed some light on this from an EU law perspective.

Therefore, this article explains what constitutes emotion data and the scope and limitations of EU law aiming to protect the persons to whom emotion data belongs. We focus on provisions in the General Data Protection Regulation (GDPR) and, to a lesser extent, on the Digital Services Act (DSA). We also discuss provisions of the AI Act proposal, which is currently undergoing the EU's trialogue process. The EU regulations we discuss have an impact on the global AC community because of their (extra)territorial scope (Article 3(2) GDPR, Article 2 DSA, Article 2 AI Act proposal). Our article also highlights some of the consequences, including harm, that processing of highly sensitive data may have for the individuals concerned. By doing so, the article aims at raising awareness among the AC community about the importance of protecting the rights of individuals even when the law struggles to keep up with technological developments that look inside our most human side.

Our article outlines that emotion data is not protected as 'special data' according to Art. 9 of the GDPR despite its sensitive nature and the related impacts processing such data may have on people. As such, it is also tricky for the affective computing community to consider the applicable legal requirements when developing AC systems that involve study participants in the EU or intended for the EU market. For instance, processing special data is prohibited under the

This article received support from the SAILS Program at Leiden University and has partly been funded by the Safe and Sound project that received funding from the EU's Horizon-ERC program (Grant Agreement No. 101076929).

GDPR unless an exception applies. Whether processing of personal data used to detect or derive emotion data falls under the framework applicable to special personal data (Art. 9 GDPR) depends on the approach taken in AC. Approaches that process physiological information fall under the scope of Article 9 GDPR, whereas visual approaches relying on the processing of facial expressions do not.

## II. EMOTIONS AND EMOTION DATA

Since ancient times, emotions have been considered significant drivers of human action and essential aspects of human decision-making processes [12]. Nevertheless, although emotions are an essential part of the human experience, the essence of emotions becomes elusive when we try to define them [13]. Despite this lack of definition, research and advances in understanding emotions and their role in human life have been made in various disciplines, such as philosophy, music, sociology, and neuroscience [14]. In psychology, for instance, emotions are discussed as particular affective states that humans experience temporarily [15].

Researchers in affective sciences have proposed several taxonomies to categorize emotions in everyday experiences [17]. The most popular taxonomy is called 'basic emotions,' which are assumed to be universally present in all humans [18]. According to this taxonomy, there are six basic emotion categories: anger, disgust, fear, happiness, sadness, and surprise [19], to which additional categories, such as anxiety, guilt, shame, pride, compassion, relief, hope, and love were added over time [20-21]. The experience of emotions and the mechanisms determining our behavior or action selection, including, e.g., facial expressions, correlate somehow. Ekman [22] developed the theory that certain facial expressions reveal universal basic emotions. Quite early, the basic emotion taxonomy was subject to substantial disagreement, especially over the extent to which the origins of facial expressions are innate or sociocultural and whether emotions could accurately be inferred from human facial expressions [18, 20-21, 23] which is still a matter of debate today [24].

Instead of elaborating on existing definitions, we will use the notion of emotion data. Emotion data is information relating to the emotions of an individual. For the sake of simplicity, emotions refer to the six basic emotion categories: anger, disgust, fear, happiness, sadness, and surprise. These emotion categories have received the lion's share of attention in scientific research [24]. This also makes sense from an affective computing perspective, as most approaches in the field rely on basic emotion categories. Since the beginning, fundamental emotion theories and their emotion categories have been the models of choice in computer science and engineering [16]. According to a review performed ten years ago, most systems were concerned with detecting the six basic emotions [17]. While we are unaware of a more recent comprehensive review, most modern AC systems also seem to focus on these emotion categories or alterations [25]. Real-world applications are Amazon's wearable 'Halo' [26], Spotify's patented voice assistant [27], and Amazon's patent enabling Alexa to recognize the user's emotional state [28].

## III. EMOTION DATA HARMS

The processing of emotional data can have significant adverse effects on people that make it legally relevant. Emotions are a private, sometimes intimate part of people's lives [29]. In particular, emotions are central to social connections – people become vulnerable, establish trust, and build relationships by revealing them to each other [30]. Emotional trust assumes the possibility of being hurt by another person [31]. Therefore, people often keep emotions private and decide with whom, when, and the contexts to share them [29]. By processing emotional data, machines provide access to information about people's emotional lives that are private and intimate [32]. Therefore, the most often discussed concern regarding the processing of emotion data is the risk of undermining "informational privacy" or people's interest in being in charge of the information about themselves, including their emotions [33]. However, the adverse effects of emotion data processing go far beyond privacy issues. Depending on how emotional data is *used*, its processing can result in manipulation and discrimination (or oppression) of individuals, which can lead to economic, relational, psychological, and physical harm to individuals, and also societal threats.

Manipulation can be defined as the hidden influence achieved by targeting to exploit people's decision-making vulnerabilities [34]. There is a scientific consensus that humans are not entirely rational in their decision-making. Behavioral science reveals that humans act on *heuristics* to decrease the complexity of real-life situations and shortcut everyday decisions [35]. This act often results in cognitive *biases* or systematic errors in judgment that the manipulator can exploit. However, emotions are another source of vulnerability to manipulation [34]. For example, in William Shakespeare's tragedy Othello, written around 1600, Iago manipulates Othello because he knows his insecurity, love, and jealousy.

The potential for manipulative use is a central concern in processing emotional data. One paradigm example of such a service can be manipulative advertising. Once a widespread science-fiction practice (e.g., Spielberg's Minority Report), using emotion data for targeting advertisements has become particularly relevant in the online advertising industry through techniques such as "dynamic emotional targeting" or "emotion analytics" [36]. Such targeting is made possible due to the increased ability of online platforms, such as Google and Meta, to identify emotional data by analyzing keyboard typing patterns [37], video data [38], and metadata [39]. Targeting advertisements based on emotion data can result in manipulation when, for example, the internet user is not aware that data about their sadness is used to target them with a video-game advertisement – the user does not know precisely how they are being influenced, making a decision that they cannot regard as their own.

From a deontological stance, manipulation is harmful because it undermines personal *autonomy* or disables a person's capability for an authentic choice [33]. However, it can also lead to various adverse consequences that are also legally relevant. For example, it can result in economic loss: subscribing to a video game after manipulative advertising would mean the user pays needlessly. This may lead to direct financial loss to a consumer, structural inefficiencies, and market failure. Moreover, manipulation can lead to time loss (e.g., a user playing a video game instead of spending time more authentically). Manipulating users based on their emotional data can also harm their psychological and physical health and integrity. This is particularly true when emotional data relates to a person's mental health. In one extreme example, an online personalization algorithm that identified 14-year-old Molly Russel as depressed targeted her with content about self-harm and suicide [40]. Eventually, Russel developed a severe depressive disorder and later ended her life. While linking such cases directly to the processing of emotion data may seem far-fetched, a coroner directly linked the personalization algorithm as the cause of her death.

On the other hand, using emotion data can exacerbate inequality or otherwise discriminate. For example, a person's temporary anxiety when considering airplane tickets can be used to extract surplus profit by charging a higher price. This can exacerbate inequality as low-income people may be more likely to experience anxiety about ticket prices. Moreover, when emotion data reveals mental health conditions, such as, for example, showing a mood disorder, this can be used in employment decisions or other decisions that may significantly affect the person's life. Finally, when emotion data is processed for public security purposes, for example, at border control, it may disadvantage historically marginalized groups and ethnic minorities who may find immigration or security lines relatively more "stressful" [41].

Furthermore, if emotion data is processed in a way that undermines information privacy, such processing may have significant risks to other interests. For example, it may have chilling effects on free speech – the feeling of being watched and classified can act as an intimidation mechanism and limit their self-expression. On the other hand, if emotion data is processed in the context of an interrogation, it may undermine an individual's interest in avoiding self-incrimination [41].

Lastly, emotions are felt as personal and express what a person cares about [42]. As people often rely on emotions to relate to their authentic core selves, processing emotion data can interfere with constructing one's selfhood. The construction of one's selfhood can be considered one of the most sensitive areas, and interfering with such a process can be regarded as treating persons as less than human, undermining their dignity, which is the foundational value in the European Union [43].

## IV. EU LAW AND EMOTION DATA

In law, expressions, and attributions of emotions historically played a critical role in legal decision-making, particularly in criminal law [44]. An accused individual's physical movements were considered to indicate inner emotions and ultimately used to determine guilt or innocence. The lie detector developed by Hugo Münsterberg is a prime example [43]. A recent example is the automated border control system called IBORDERCTRL [8]. This research project, which the EU funded, analyses travelers' micro-gestures to determine if the interviewee is lying.

There are existing and emerging areas of EU law in which emotion data plays an important role. We discuss the GDPR and the DSA. The former is active for five years and shall apply as of 24 February 2024. We also consider the AI Act proposal, which is subject to the EU's legislative procedure at the time of writing.

### A. The EU General Data Protection Regulation

The GDPR only applies to the processing of personal data. Personal data is defined as a concept with four elements: i) any information ii) relating to iii) an identified or identifiable iv) natural person (Article 4 GDPR). Although emotions are felt as personal because they are related to a person's values [29] and express what a person cares about [42], such information is not per se considered personal data from a legal perspective [43], as confirmed by the examples of two European Supervisory Authorities. Processing emotion data does not fall under the material scope of the GDPR in case the individual concerned is neither identified nor identifiable. An example of this may be found in a billboard installed at Piccadilly Circus in London using AC to broadcast ads based on people's age, gender, and mood [46]. The same holds when retailers capture data about age, gender and observed emotions of retail customers without identifying them. Providers usually argue that the system only stores anonymous data like age and gender. Therefore, the captured emotion data "belong to no one" because they cannot be linked to individuals [47]. Some EU Data Protection Supervisory Authorities seem to agree [48], [49]. Usually, however, the use of emotion data amounts to the processing of personal data because individuals are identified or identifiable, i.e., if AC systems are used in a hiring context or by call center agents.

*Emotion data as special data*

In most cases, emotion data constitutes personal data. The question arises whether such data is specifically protected due to its sensitive nature. The answer to this question is complicated. First and foremost, emotion data is not specifically protected under Article 9 GDPR. This provision regulates the processing of special categories of personal data ('special data'). It prohibits "*the processing of personal data revealing racial or ethnic origin, political opinions, religious or philosophical beliefs, or trade union membership, and the processing of genetic data, biometric data for the purpose of uniquely identifying a natural person, data concerning health or data concerning a natural person's sex life or sexual orientation.*" Processing such data is prohibited and only allowed if an exception in Article 9(2) GDPR applies.

Because emotion data is not listed in this exhaustive enumeration of special data, it is never specifically protected as such [50-51]. Nonetheless, the data used to detect emotion data may constitute special data. Ultimately, the *approach taken in AC* determines whether processing *personal data used to detect or derive* emotion data falls under the scope of Article 9 GDPR. A current survey distinguishes between AC's single-modal and multi-modal affect recognition approaches [52]. Single-modal approaches are divided into text sentiment analysis, audio emotion recognition, visual emotion recognition focusing on facial expression and body gestures, and physiological-based emotion recognition systems [52]. The latter include AC systems that detect emotional states from electroencephalograms (EEGs) and electrocardiograms (ECGs). ECG-based emotion recognition systems record the physiological changes of the human heart in order to detect the corresponding waveform transformation, which provides information for emotion recognition [52]. For instance, Hsu et al. presented an ECG-based emotion recognition system for music listening [53]. EEG is a non-invasive method that detects and registers electrical activity in the brain [54]. EEG-based emotion recognition systems directly measure changes in brain activities, which provide internal features of an emotional state [52], [55].

Merely physiological-based emotion recognition systems in AC involve processing special data as defined in the GDPR. Information processed by these systems falls under the definition of health data, which not only covers physical or mental health but also "any information (…) on the *physiological* or biomedical state of the data subject independent of its source" (Recital 35). Consider, for instance, AC systems that infer emotion data from physiological data such as heart rate, blood pressure, and skin conductance. Such information falls under the definition of health data and is protected as a special category of personal data according to Article 9 of the GDPR.

Conversely, most of the single-modal affect recognition systems pursued in AC do *not* amount to the processing of special data. AC systems deploying approaches such as text sentiment analysis, audio emotion recognition, and visual recognition of emotion focusing on facial expressions and body gestures do *not* directly involve processing special categories of personal data (recitals 51-53). Information processed within these approaches and derived emotion data are thus not protected as special personal data under the GDPR, despite their sensitive and intimate nature. This also holds when biometric data is used for AC to detect the emotional state of the individual concerned. Think about automated face analysis (AFA) approaches relying on the facial action coding system (FACS) [56], [57]. Within these approaches, biometric data in facial expressions are *not* processed to identify an individual. The same applies to AC systems aiming to detect emotion data from an individual's voice and speech. According to the wording of Article 9(1) GDPR, biometric data is only protected as special personal data if it is used *for the purpose of uniquely identifying* an individual. This means "processed through a specific technical means allowing the unique identification or authentication of a natural person" (recital 51 GDPR).

According to regulatory guidance, biometric identification typically involves "the process of comparing biometric data of an individual (acquired at the time of the identification) to several other biometric templates stored in a data database (i.e., a one-to-many matching process)" (WP 193, 27 April 2012). For example, HumeAI, or formerly HireVue, provides AC-powered tools to help recruiters to assess personality traits and detect emotional states of job candidates disclosed during automated video assessments based on facial expressions. These systems do not process biometric data in the form of facial expressions for the purpose of uniquely identifying the job candidate as required by Article 9(1) GDPR. Instead, they detect a candidate's emotional state during the automated video assessment. Identification is achieved through other means, namely when the candidate reveals their name or other identifiable information. Likewise, according to the automated voice analysis, an AC system that advises the call agent to speak with more empathy because the customer seems to be angry does not process biometric data for identification purposes.

In conclusion, highly sensitive emotion data never constitutes special personal data protected explicitly in the GDPR. In some cases, AC systems process special data to *derive or detect* emotion data. This applies to physiological-based emotion recognition systems that process information like heart rate, blood pressure, and skin conductance. Such information constitutes health data and is protected as a special category of personal data in the GDPR. Nonetheless, the highly sensitive detected emotion data never constitutes special data under the GDPR, irrespective of which affect recognition (single-modal or multi-modal) approach in AC is deployed. Inherently sensitive personal data is not specifically protected in EU data protection law. This leads to a significant gap in legal protection. It could be argued that emotions should be regulated like human speech or text because both somehow define humanity. In our view, the inherently highly sensitive nature of emotion data and the close link with one's personhood merits specific protection.

*Data protection principles*

Article 5 of the GDPR stipulates the principles governing any personal data processing. These principles provide the basis for protecting personal data in EU data protection law [58]. In the context of AC, three EU data protection law principles are particularly relevant. These are the fairness and transparency principles enshrined in Article 5(1) lit a GDPR as well as the accuracy principle according to Article 5(1)(d) GDPR. Moreover, according to the principle of data protection by design and default (Article 25 GDPR), these principles must be considered when developing and designing AI systems that process personal data.

*Fairness*

Even though the fairness principle is a crucial tenet of EU data protection law, its role has thus far endured being elusive [59] due to the lack of judicial guidance. Nonetheless, regulatory guidance (Guidelines 2/2019) and regulatory enforcement at the EU level (in the form of binding decisions adopted by the EDPB according to Article 65 of the GDPR) identify five key elements of the fairness principle. These elements are the autonomy of data subjects concerning data processing, their reasonable expectations, ensuring power balance between controllers and data subjects, avoidance of deception as well as possible adverse consequences of processing, and ensuring ethical and truthful processing (EDPB binding decisions 3/2022, 4/2022, 5/2022). Fairness is an *overarching* principle beyond transparency [60].

Processing personal data occurring in the context of AC raises the question of whether such processing complies with the fairness principle. Processing emotion data enabled by AC could be misleading, mainly because the accuracy of AC has been severely questioned [24], [61-62]. Processing emotion data through AC could be detrimental and unexpected for the individuals concerned. Imagine an employer that uses automated video assessments provided by HumeAI and formerly HireVue to detect the emotional states of candidates shown during these assessments. Particularly in these circumstances, processing emotion data employing AC may have adverse consequences for the data subject. Perhaps for this reason, HireVue halted the operation of its services' component analyzing the facial expressions of applicants [63].

Considering the sometimes questionable accuracy of AC (see the accuracy principle), the ubiquitous manner of processing (see the transparency principle), the sensitive nature of the personal data processed, and possible adverse effects for the candidate, it seems reasonable to conclude that such processing does not comply with the fairness principle. The power asymmetry between the employer and the applicants also plays a role. Fairness aims to balance precisely these kinds of power asymmetries and to prevent adverse effects in concrete circumstances [64]. Here, the adverse effects are apparent. Rather sensitive personal data is processed to determine whether the applicant will receive a job offer. Undoubtedly, this decision has a considerable effect on the candidate. Even so, relying on a human observer instead of an emotion recognition system may lead to similar problems. Moreover, processing emotion data can also improve fairness, for instance, by highlighting human biases.

Of course, there are also examples outside the HR domain. Think about the automated border control system called IBORDERCTRL, which tries to determine if the interviewee is lying. It may be severely questionable whether such processing is fair. The use of AC might also be unfair within other domains, for instance, when implemented in cars, classroom teaching aids, smart toys, virtual assistants, and targeted advertisements. Although AC systems are predominantly developed in the West, they are sold to global marketplaces. Algorithms are hardly modified for racial, cultural, ethnic, or gender differences [65-66]. Importantly, AC systems may violate the fairness principle when emotion data is used to manipulate data subjects and adversely affect their personal autonomy.

*Transparency*

The transparency principle inherent in Article 6(1) GDPR requires that it must be transparent to natural persons "that personal data concerning them are collected, used, consulted or otherwise processed" (recital 39). It is further specified in articles 12 to 14 of the GDPR in the form of data controller obligations to provide certain information to the data subject, for instance, about the purposes of processing.

As propagated by Picard [32], transparent processing performed by AC systems presupposes that individuals know what emotions the machine has recognized. However, this does not seem to be the case when the transparency obligations contained in the GDPR are applied in practice. Usually, emotion data are not directly provided by the data subject concerned. Instead, emotion data is *inferred* by AC systems based on personal data collected from data subjects. This is important because Article 13 of the GDPR only applies when personal data are obtained or observed from the data subject. Because complex computing is needed to detect emotion data, it cannot be considered as simply observed data. Thus, the transparency requirements enshrined in Article 13 of the GDPR are not triggered [67]. Emotion data must be considered as being *inferred* from personal data provided by the data subject. Regarding inferred personal data, regulatory guidance states that the controller must provide information about the intended *purpose* and the *categories* of the inferred data (WP 260 rev.01, 11 April 2018).

When this guidance is applied to processing the AC software provided by HumeAI and formerly HireVue, the employer merely needs to inform the candidate about the intended purpose of creating inferred data and the category of inferred data at the commencement phase of the processing cycle. Hence, the prospective employer is not obliged to inform the candidate about the specific emotions detected by the system. The same holds if the prospective employer receives emotion data from an independent assessment provider (i.e., not obtained from the data subject). Article 14 GDPR, which is applicable in this scenario, simply requires the employer to inform the candidate about the category of personal data (e.g., 'emotion data'). The candidate will not know what specific emotions the system detected. The same conclusion can be drawn for Speech Emotion Recognition (SER) applications used by call centers to detect emotion data of customers employing automated speech analysis.

This leads to a significant loophole [50]. Candidates and callers do not know what specific emotions are being detected about them. We use the term 'loophole' because we share Picard's view that transparent processing presupposes that individuals can see what emotion the machine recognized [32]. However, as outlined in this section, EU data protection law does not oblige controllers to disclose the specific emotion data detected by the AC system.

*Accuracy*

Article 5(1)(d) GDPR states that the processing of personal data must be accurate. The accuracy principle protects the individual concerned from being irrationally or unfairly treated based on wrong and inaccurate representations [68]. According to regulatory guidance, accurate means "accurate as to a matter of fact" (WP 225, 26 November 2014). Case law of the CJEU (Case C-434/16) indicates that the precision required for processing personal data is determined by the purpose of the processing [68]. Thus, the assessment as to whether personal data is accurate and complete depends on the *purpose* for which it was collected.

Personal data generated using AC are subject to this accuracy principle. However, different studies have challenged the idea that a person's emotional state can accurately be inferred from his or her facial movements. An extensive study suggests that facial movements are not diagnostic displays that reliably and precisely signal particular emotional states regardless of context, person, and culture [24]. Another study revealed that the accuracy levels of eight commercial automatic classifiers used for facial affect recognition were consistently lower when applied to spontaneous affective behaviors than "posed" affective behavior. Recognition accuracy rates of the tested classifiers varied from 48% to 62% [69].

Additionally, the accuracy of emotion data inferred by other means, such as physiological data or speech, has been questioned [70]. Research that explores emotional aspects of speech has been restricted to laboratory conditions with noise-free, uncompressed, and full-bandwidth audio recordings. Recent studies, however, indicate that speech compression, filtering, band reduction, and the addition of noise significantly reduce accuracy [71]. Nevertheless, speech emotion recognition is already being applied 'in the wild', and emotions are inferred from speech recorded or streamed in natural and daily-life environments, likely with significantly lower accuracy rates. Also, Recital 26c of the AI Act proposal discussed in Section B below mentions the following 'shortcomings' of current AC systems: limited reliability, lack of specificity, and limited generalizability. Thus, the processing of personal data by means of AC creates severe tension with the accuracy principle. If companies act upon such arguably inaccurate data and treat individuals in a particular manner, it could lead to severe problems for the individuals concerned.

*B. AI Act proposal*

In 2021, the EU Commission proposed the AI Act [72]. After multiple amendments, the European Parliament has adopted its negotiation position for the AI Act [73] ('AI Act proposal'), which now undergoes the EU's trialogue process. The AI Act proposal covers aspects ranging from product safety law to fundamental rights. It contains an exemption for scientific research. Research, testing, and development activities of AI systems occurring before putting such systems on the market do not fall under the AI Act, except for testing in real-world conditions (Article 5d).

Article 3 (34) of the AI Act proposal directly relates to AC systems. It defines 'emotion recognition systems' as systems used "for the purpose of identifying or inferring emotions, thoughts, states of mind or intentions of individuals or groups on the basis of their biometric and biometric-based data." Biometric-based data are defined in Article 3 (33a) as "data resulting from specific technical processing relating to physical, physiological or behavioral signals of a natural person." Recital 26c of the AI Act proposal names facial expressions, movements, pulse frequency, and voice as examples. Thus, the definition of emotion recognition systems ('ERS') is very broad, arguably covering both single-modal and multi-modal affect recognition approaches in AC as introduced in the previously mentioned survey [52].

The AI Act proposal takes a risk-based approach toward emotion recognition systems ('ERS'). Firstly, Article 5(1) dc prohibits ERS from being used in law enforcement, border management, the workplace, and education. Secondly, ERS are generally considered high-risk systems as outlined in Article 6(2) and Annex III. This means that ERS must meet specific compliance requirements, such as those pertaining to risk management systems, accuracy, data governance, transparency, human oversight, specific technical documentation, and record-keeping (Articles 9-15). Deployers of AI systems must also perform a fundamental rights impact assessment (Article 29a). We focus on transparency and accuracy requirements relating to ERS.

*Transparency*

Article 52(2a) AI Act proposal obliges entities under whose authority the ERS is used ('deployer') to inform individuals concerned about *the operation of the system*. Accompanying recital 70 explains that "natural persons should be notified" when exposed to ERS. None of the recitals further clarify what information about the system's operation precisely entails. Arguably, it simply means to make natural persons aware that they are exposed to an ERS. Hence, deployers of ERS are not obliged to inform individuals about what specific emotion the system detected. Similar to the situation with the GDPR, this contradicts what Picard propagated: individuals should be able to know what emotion the machine recognizes [32]. In conclusion, the AI Act does not fill the current loophole in EU data protection law. Under both the GDPR and the AI Act proposal, individuals do not know what specific emotions are being detected about them.

*Accuracy*

Under the AI Act proposal, ERS are high-risk AI systems according to Article 6(2) and Annex III. Article 15(1) of the AI Act proposal requires that high-risk AI systems are designed and developed in such a way that they achieve, in light of their intended purpose, an "appropriate level of accuracy" Levels of accuracy and relevant accuracy metrics must be declared in the accompanying documentation. This contains, inter alia, detailed information about the AI system's degree of accuracy for specific *persons* or *groups of persons* on which the system is intended to be used. The documentation must also disclose the overall expected level of accuracy concerning its intended purpose. The latter resembles the accuracy principle in data protection law as discussed under A) above. Nonetheless, accuracy under the AI Act appears to be much more specific, as the degree of accuracy must be disclosed regarding specific persons or groups of persons on which the ERS will be used.

*C. Digital Services Act*

The newly enacted Digital Services Act (DSA) updates the EU's legal framework for intermediary services, including information society services and online platforms. Certain provisions are only applicable to very large online platforms (VLOP) and very large online search engines (VLOSE), as defined by the legislation (Art. 33 DSA).

Several provisions in the DSA may impact AC systems, many of which relate to advertisements or recommender systems. Notably, the DSA introduces a prohibition of advertising to minors based on their profiling (Art. 28(2) DSA), which inherently includes profiling that uses emotion data. Advertising based on profiling that uses special data is prohibited (Art. 26(3) DSA). However, as seen above, emotion data does not necessarily constitute special data, and thus may fall outside the scope of this provision. For entities that are designated as VLOP or VLOSE, the DSA includes requirements for risk assessments and the related mitigation of risks (Art. 34, 35 DSA), yearly independent audits (Art. 37 DSA), and the provision of a recommender system that is not based on profiling (Art. 38 DSA).

The risk assessments are to include systemic risks to "any actual or foreseeable negative effects for the exercise of fundamental rights," particularly those to human dignity, respect for private and family life, the protection of personal data, non-discrimination, and a high level of consumer protection, among others (Art. 34(1)(b) DSA). They are also to include systemic risks to "any actual or foreseeable negative effects in relation to… serious negative consequences to the person's physical and mental well-being" (Art. 34(1)(d) DSA). Given the implications of the use of emotion data on individuals' autonomy, privacy, and mental well-being (*see* Section III), using such data in services provided by VLOP or VLOSE should form part of their risk assessment.

They are then obligated to "put in place reasonable, proportionate and effective mitigation measures, tailored to the specific systemic risks identified" (Art. 35(1) DSA). This can include "adapting the design, features or functioning of their services," adapting their algorithmic systems (including recommender systems), or adapting their advertising systems (Art. 35(1)(a), (d), and (e) DSA).

Finally, VLOP or VLOSE entities must "provide at least one option for each of their recommender systems which is not based on profiling" (Art. 38 DSA). This is important for individuals as they can prevent recommender systems from using their emotion data. Interestingly, the DSA does not require such an option for advertisements.

The DSA also introduces new requirements regarding transparency not covered by the GDPR and AI Act proposal. Online platforms that present advertisements must provide individuals with information "about the main parameters used to determine the recipient to whom the advertisement is presented and, where applicable, about how to change those parameters" (Art. 26(1)(d) DSA). Online platforms must also disclose "the main parameters used in their recommender systems, as well as any options for the recipients of the service to modify or influence" them (Art. 27(1) DSA). This must include the most significant criteria used in determining the information suggested to individuals and the reasons for their importance (Art. 27(2)(a) and (b) DSA). Where AC systems are used, both Articles require online platforms to mention emotion data when it is used as a primary parameter for advertisements or in recommender systems. However, this leaves a gap when emotion data may be used as a secondary parameter, as its use may not be disclosed. Moreover, even when emotion data may be a main parameter, online platforms are unlikely to detail which emotion the AC system recognized [32]. Given the novelty of these provisions, it remains to be seen how online platforms will attempt to comply with them or how supervisory authorities will enforce them.

## V. Conclusions

In this article, we have outlined that emotion data does not constitute special data in EU data protection law despite its sensitive nature. The GDPR fails to keep up with technological developments, which leads to a lacuna of protection. As such, it is also tricky for the affective computing community to consider the GDPR's legal requirements when developing AC systems because of its lack of clarity and knowledge-specific sector. Whether the processing of personal data used to detect or derive emotion data falls under the scope of special personal data, according to Art. 9 of the GDPR, depends on the *approach* taken in affective computing. Approaches that process physiological information do, whereas visual approaches that rely on biometric data (e.g., facial expressions) do not—at least not directly. The AI Act proposal introduces yet another term relevant to AC: biometric-based data. This is data resulting from specific technical processing relating to a natural person's physical, physiological, or behavioral signals. These legal nuances make it difficult for the AC community to sufficiently consider the applicable legal requirements when developing AC systems intended for the EU market or when working with study participants in the EU.

Processing emotion data by means of affective computing systems may be detrimental to critical elements of the fairness principle contained in the GDPR. Also, the processing of emotion data creates severe tensions with the accuracy principle enshrined in the GDPR. Several studies have questioned the accuracy of emotion data inferred by means of AC [24], [69-71]. Moreover, EU data protection law does not oblige deployers of AC systems to inform individuals about the specific emotions detected by the system, contrary to what Picard propagates [32]. It seems that the AI Act, for now, will not fill this loophole. Moreover, the AI Act may limit specific harmful uses of emotion data. This can mainly be the case regarding manipulation that leads to psychological or physical harm. However, arguing that the risk of such harm exists will be a difficult exercise. Furthermore, AI Act leaves unaddressed other harms (e.g., time loss, economic loss) of manipulative uses of emotion data processing.

The DSA introduces several new obligations for online platforms relevant to AC systems and emotion data. These include risk assessments and mitigation measures, independent audits, and the implementation of audit recommendations, all accompanied by reports that must be made public. In addition, transparency requirements regarding recommender systems and advertisements may additionally shed light on how AC systems and emotion data are used in practice, though there are some limitations. Further, the DSA introduces several provisions that limit how emotion data may be used. For instance, both the use of special data used for profiling in advertisements, as well as the profiling of minors for advertisements, is prohibited.

## Ethical Impact Statement

Compliance with legislation cited in our article does not mean that all ethical concerns are satisfied. This holds particularly true when considering the gaps of legal protection we have identified. These loopholes shall not be exploited. Furthermore, ethical concerns outlined in Section III should be taken into account.

## Acknowledgment

The authors would like to give a special thank you to Joost Batenburg, the coordinator of SAILS Program, a Leiden University wide initiative aiming to facilitate collaboration across disciplines on the use of AI.